%% file: main.tex
\documentclass[conference]{IEEEtran}
\IEEEoverridecommandlockouts

\usepackage{cite}
\usepackage{amsmath,amssymb,amsfonts}
\usepackage{algorithmic}
\usepackage{graphicx}
\usepackage{textcomp}
\usepackage{xcolor}
\def\BibTeX{{\rm B\kern-.05em{\sc i\kern-.025em b}\kern-.08em
    T\kern-.1667em\lower.7ex\hbox{E}\kern-.125emX}}

\usepackage{comment}
\usepackage{times}
\usepackage{helvet}
\usepackage{courier}
\usepackage{algorithm}
\usepackage{algorithmic}
\usepackage{adjustbox,makecell}
\usepackage{newtxmath}
\usepackage{bm}
\usepackage{url}
\usepackage{pifont}
\usepackage{booktabs}

\definecolor{mygray}{gray}{0.9}

\begin{document}

\newcommand{\name}{HDLxGraph}

\title{{\name}: Bridging Large Language Models and HDL Repositories via HDL Graph Databases
}

\author{\IEEEauthorblockN{Pingqing Zheng}
\IEEEauthorblockA{\textit{University of Minnesota, Twin Cities} \\
Minneapolis, MN, USA \\
pingqingzheng13@gmail.com}
\and
\IEEEauthorblockN{Jiayin Qin}
\IEEEauthorblockA{\textit{University of Minnesota, Twin Cities} \\
Minneapolis, MN, USA \\
qin00162@umn.edu}
\and
\IEEEauthorblockN{Fuqi Zhang}
\IEEEauthorblockA{\textit{University of Minnesota, Twin Cities} \\
Minneapolis, MN, USA \\
zhan7076@umn.edu}
\and
\IEEEauthorblockN{Niraj Chitla}
\IEEEauthorblockA{\textit{University of Minnesota, Twin Cities} \\
Minneapolis, MN, USA \\
chitl013@umn.edu}
\and
\IEEEauthorblockN{Zishen Wan}
\IEEEauthorblockA{\textit{Georgia Institute of Technology} \\
Atlanta, GA, USA \\
zishenwan@gatech.edu}
\and
\IEEEauthorblockN{Shang Wu}
\IEEEauthorblockA{\textit{Northwestern University} \\
Evanston, USA \\
swu@u.northwestern.edu}
\and
\IEEEauthorblockN{Yu (Kevin) Cao}
\IEEEauthorblockA{\textit{University of Minnesota, Twin Cities} \\
Minneapolis, MN, USA \\
yucao@umn.edu}
\and
\IEEEauthorblockN{Caiwen Ding}
\IEEEauthorblockA{\textit{University of Minnesota, Twin Cities} \\
Minneapolis, MN, USA \\
dingc@umn.edu}
\and
\IEEEauthorblockN{Yang (Katie) Zhao}
\IEEEauthorblockA{\textit{University of Minnesota, Twin Cities} \\
Minneapolis, MN, USA \\
yangzhao@umn.edu}
}
\maketitle

\input{Sections/0-Abstract}

\input{Sections/1-Introduction}
\input{Sections/2-Related_Work}

\input{Sections/3-Framework}

\input{Sections/4-Evaluation}
\input{Sections/5-Conclusion}
\bibliographystyle{unsrt}
\bibliography{references,generation}

\clearpage

\end{document}

%% file: Sections/0-Abstract.tex
\begin{abstract}
Retrieval Augmented Generation (RAG) is an essential agent for Large Language Model (LLM) aided Description Language (HDL) tasks, addressing the challenges of limited training data and prohibitively long prompts.
However, its performance in handling ambiguous queries and real-world, repository-level HDL projects containing thousands or even tens of thousands of code lines remains limited.
Our analysis demonstrates two fundamental mismatches, structural and vocabulary, between conventional semantic similarity-based RAGs and HDL codes. 
To this end, we propose {\name}, the first framework that integrates the inherent graph characteristics of HDLs with RAGs for LLM-assisted tasks. 
Specifically, {\name} incorporates Abstract Syntax Trees (ASTs) to capture HDLs’ hierarchical structures and Data Flow Graphs (DFGs) to address the vocabulary mismatch.
In addition, to overcome the lack of comprehensive HDL search benchmarks, we introduce {HDLSearch}, an LLM-generated dataset derived from real-world, repository-level HDL projects.
Evaluations show that {\name} improves search, debugging, and completion accuracy by 12.04\%/12.22\%/5.04\% and by 11.59\%/8.18\%/4.07\% over state-of-the-art similarity-based RAG and software-code Graph RAG baselines, respectively. The code of {\name} and HDLSearch benchmark are available at \url{https://github.com/UMN-ZhaoLab/HDLxGraph}.

\end{abstract}

%% file: Sections/1-Introduction.tex
\section{Introduction}
\label{sec:intro}

Recent advances in Large Language Models (LLMs) for software language understanding and generation \cite{wang-etal-2023-codet5,lozhkov2024starcoder2stackv2} have inspired efforts to extend their capabilities to facilitate Hardware Description Language (HDL) code designs. Prior works have demonstrated LLMs’ potential in generating~\cite{zhao2024codevempoweringllmsverilog,thakur2023verigenlargelanguagemodel,10.1145/3649329.3658493} and debugging HDL code~\cite{yao2024hdldebuggerstreamlininghdldebugging,10691824,pu2024customizedretrievalaugmentedgeneration}. 
However, LLM performance in HDL-related tasks remains hindered by limited training data and degradation caused by long prompts.
To address these issues, researchers have integrated \textit{Retrieval-Augmented Generation} \textit{(RAG)}, which retrieves relevant HDL fragments from high-quality HDL codes to supplement knowledge gaps and reduce prompt length~\cite{pu2024customizedretrievalaugmentedgeneration,gao2024autovcodersystematicframeworkautomated}.

\begin{figure}[t]
     \centering
    \includegraphics[width=1.0\linewidth]{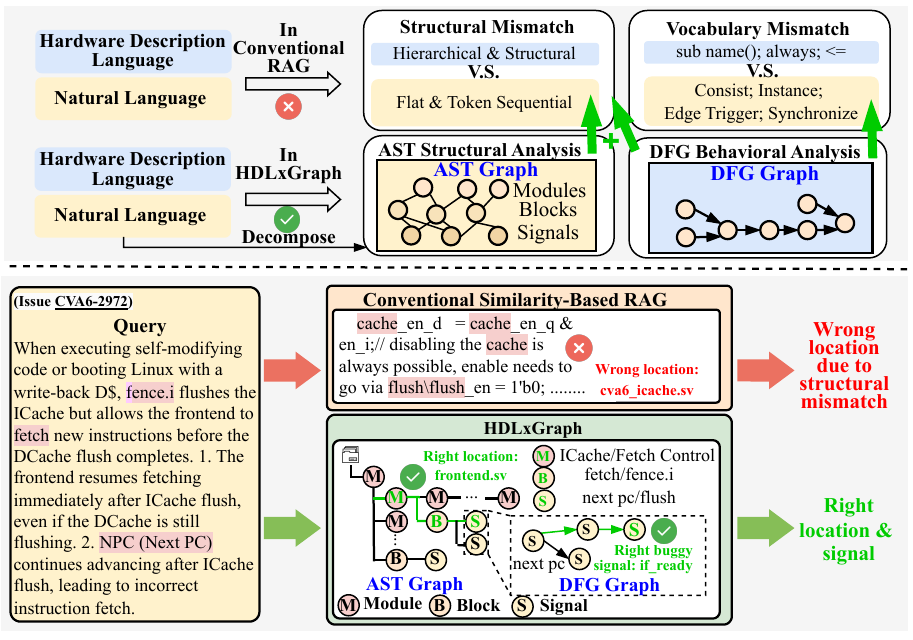} 
    \vspace{-2em}
    \caption{(Top) An illustration of two fundamental mismatch between HDL and natural language in conventional RAG, including \textit{structural} and \textit{vocabulary} mismatches. And (Bottom) a demonstration of {\name}'s efficiency in bridging these mismatches by incorporating graph information, using an HDL debugging example for a CVA6 RISC-V HDL implementation~\cite{cva6}. }
    \label{fig:similarity-search_limitations_fig0}
\end{figure}

Despite their potential, existing RAG approaches for HDL primarily rely on semantic similarity-based retrieval, which yield low recall when handling intricate and ambiguous queries or large and complex HDL repositories with thousands or even tens of thousands of code lines. 
Through analyzing semantic similarity-based RAGs and HDL codes, we find \textbf{two fundamental mismatches} between them (as illustrated in Figure~\ref{fig:similarity-search_limitations_fig0} (top)). 
First, natural language queries are flat and sequential while HDL designs have a hierarchical structure with multi-level entities, i.e., modules, blocks, and signals. We summarize this as the \textit{structural mismatch}. In addition, cross-file and cross-module analysis is required to capture this hierarchical structure. 
Second, domain-specific HDL terminologies, such as the operators and key words, differ from their natural language descriptions. This \textit{vocabulary mismatch} further prevents conventional RAGs from effectively learning HDL code semantics. 

Inspired by recent advances in \textit{Graph Retrieval-Augmented Generation} (Graph RAG)~\cite{edge2024localglobalgraphrag,guo2024lightragsimplefastretrievalaugmented} and the inherent graphic characteristics of HDL codes, we propose integrating HDL-specific graph structures into HDL-specific RAGs to address the aforementioned mismatches. 
To this end, we introduce {\name}, a novel hybrid graph-enhanced RAG framework, which incorporates two HDL-specific graphs: Abstract Syntax Trees (ASTs) and Data Flow Graphs (DFGs). 
Using ASTs, we parse a HDL repository, even with tens of thousands of code lines,
into a \textbf{code graph view} containing multi-level entity hierarchy, including modules, blocks and signals. 
By abstracting DFGs from the signal-level connections in the ASTs, we further construct a precise \textbf{hardware signal graph view} representing the signal-level data flow within the hardware design.

We use an HDL debugging example for a CVA6 RISC-V implementation, which contains over 30 modules~\cite{cva6}, to demonstrate the advantages of {\name} (shown in Figure~\ref{fig:similarity-search_limitations_fig0} (bottom)). Conventional similarity-based RAGs often mislead retrieval to \texttt{cva6\_icache.sv} due to the frequent presence of keywords containing ``Cache'' in the natural language query. In contrast, {\name} can address this issue through AST- and DGS-integrated retrieval: 
1) The AST explicitly reveals the hierarchical hardware structure. When analyzing natural language queries, {\name} can align the flat and sequential descriptions with the hierarchical HDL structure, even if the keywords indicating hierarchy appear only once or out of structural order. Consequently, {\name} accurately locates the relevant file \texttt{frontend.sv}.
2) To identify and correct the bug, {\name} then traces through the DFG from the queried buggy signal description to the actual buggy signal, \texttt{if\_ready} in this example, and guides the HDL correction for debugging.

Besides developing {\name}, we also identify the lack of comprehensive HDL code search benchmarks that include question–answer pairs derived from repository-level HDL projects.
To address this gap, we further extend {\name} to construct a new benchmark, dubbed {HDLSearch}.

In summary, our key contributions are as follows:

\begin{itemize}
    \item We propose {\name}, a novel RAG framework that integrates HDL-specific ASTs and DFGs to address the inherent mismatches between conventional semantic similarity-based RAGs and HDL codes. To the best of our knowledge, {\name} is \textbf{the first} framework to integrate HDLs' inherent graph structures with RAGs.
    \item {\name} is a hybrid graph RAG framework with enhanced understanding on HDLs' hierarchical structures and hardware data flows. Specifically, {\name} aligns flat and sequential natural language queries across hierarchical structures containing multi-level entities in the ASTs. Building on this hierarchical alignment, {\name} overcomes the vocabulary mismatch and traces through the DFGs to locate signals for downstream dubugging and completion tasks.    
    \item Based on {\name}, we further construct a new LLM-generated dataset for HDL code search, called HDLSearch, which derives query benchmark from real-world repository-level HDL projects, to address the gap of insufficient search datasets for HDL codes. 
    \item By integrating {\name} with three commonly-used LLMs of different scales and coding abilities, we demonstrate its adaptability on code search, debugging, and completion tasks. {\name} improves the accuracy of search, debugging, and completion tasks by 12.04\%/12.22\%/5.04\% and 11.59\%/8.18\%/4.07\% over state-of-the-art similarity-based RAG and software-code Graph RAG, respectively. 
\end{itemize}

%% file: Sections/2-Related_Work.tex
\section{Preliminaries}
\label{sec:rw}
\subsection{LLM-aided HDL Tasks}


\textbf{Generation.} Although LLMs excel at generating simple HDL designs, they still struggle with complex repository-level chip designs~\cite{zhang2024mgverilog, thakur2023benchmarking,liu2024rtlcoder,GPT4AIGCHip}. For example, state-of-the-art (SOTA) works~\cite{GPT4AIGCHip,gao2024autovcodersystematicframeworkautomated,tsai2023rtlfixer} often rely on predefined templates or customized RAG datasets created by human experts, where LLMs primarily fill in fixed reference code segments rather than performing general HDL generation.


\textbf{Debugging.} 
Though LLMs perform well on simple debugging tasks, they fail to achieve high coverage in more complex fixing tasks~\cite{HDL_Bug,meic,LLM4SecHW,Bug_fix}. The study in~\cite{HDL_Bug} integrates LLMs with RAG to patch functional HDL bugs. However, it still depends on manually defined error types, reflecting the limitations of current approaches in handling real-world, complex bug fixing.



\textbf{Search.} Precise code search serves as the foundation for both HDL generation and debugging. Although no prior work has directly targeted HDL search, recent studies have explored LLMs’ potential in HDL summarization~\cite{zhao2024codevempoweringllmsverilog}, a preliminary step toward effective HDL search.
However, these approaches overlook the inherent hierarchical structure of HDL, limiting their applicability to precise code search tasks.


\subsection{Graph Retrieval-Augmented Generation for Code Tasks}

Graph RAG leverages structured knowledge graphs to enhance the complex reasoning and context-awareness capabilities of RAG, and has demonstrated promising performance in various software code tasks~\cite{edge2024localglobalgraphrag}. However, despite syntactic similarities between HDLs and software languages, HDLs exhibit fundamental differ in abstraction hierarchy and behavioral modeling~\cite{du2024codegragbridginggapnatural,liu2024codexgraphbridginglargelanguage,abdelaziz2021graph4code}, rendering existing software-code  Graph RAG suboptimal for HDL-related tasks.

First, at the abstraction level, HDLs are specification languages rather than general-purpose programming languages. To facilitate circuit optimization, HDL designs typically require explicit control over hardware components, with the module serving as the primary design unit. In contrast, software languages offer multi-level abstractions (e.g., functions, classes, packages). Second, in terms of behavioral modeling, HDLs construct concurrently executed blocks using always blocks and model combinational signal propagation using assign statements (using Verilog as an example), both of which diverge from software's sequential control flow. As a result, Graph RAGs tailored for software fail to capture the true structure and semantics present in HDL code.

To integrate HDL-specific abstraction structures and behavioral patterns, we incorporate HDL-dedicated graph characteristics and customize the retrieval process accordingly to enable precise HDL queries.

\subsection{Benchmarks/Datasets for LLM-aided Tasks}
For HDL code generation benchmarks, RTLLM~\cite{lu2024rtllm} consists of 30 designs; and VerilogEval~\cite{liu2023verilogeval} presents an evaluation dataset consisting of 156 problems from HDLBits. For HDL code debugging benchmarks, LLM4SecHW \cite{LLM4SecHW} contains a repository-level bug localization and repair sets from the version control data in Github; RTLFixer \cite{tsai2023rtlfixer} introduces a Verilog syntax debugging dataset, derived from VerilogEval~\cite{liu2023verilogeval}; and CirFix \cite{cirfix} includes a repair benchmark with testbenches. 

No existing benchmark has been established for the HDL search, which is an essential step for downstream tasks such as generation and debugging. Therefore, we propose HDLSearch, the first benchmark for HDL code search, which derives query benchmark from real-world repository-level HDL projects.

%% file: Sections/3-Framework.tex
\section{Methodology}

 \begin{figure}[t]
     \centering
\includegraphics[width=1.0\linewidth]{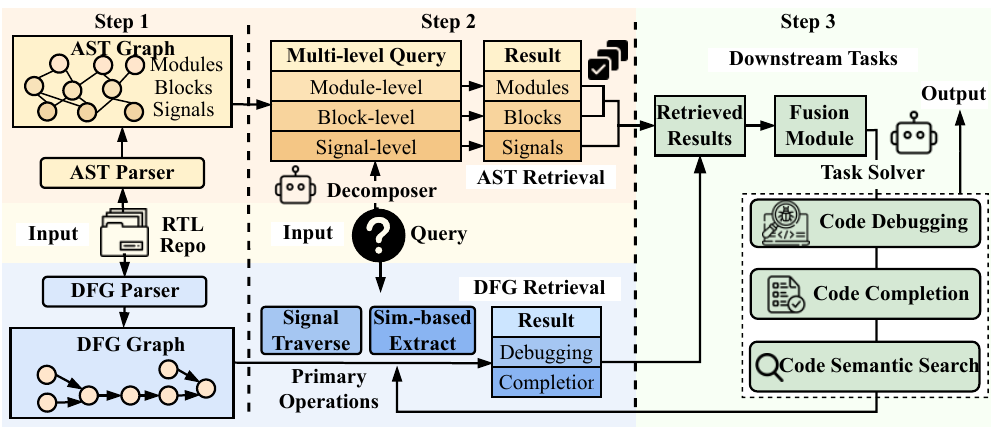} 
    \vspace{-2em}
    \caption{The overview of our proposed HDLxGraph framework.}
    \label{fig1}
\end{figure}

Figure~\ref{fig1} illustrates the overview and workflow of our proposed HDLxGraph framework, which consists of three steps: 1) Graph Database Preparation, 2) Multi-level Retrieval, and 3) Downstream Task Completion. Beginning with \textbf{Step 1}, we extract ASTs and DFGs from the input code repositories through the AST and DFG parsers, then store HDL entities and relationships as nodes and edges in a graph database (see Section~\ref{sub:build graph database}). In \textbf{Step 2} (see Section~\ref{sub:code search}), HDLxGraph utilizes an LLM called Decomposer in AST retrieval to extract the input query into structural levels, which are later sent to pre-defined searching paradigms to retrieve relevant fine-grained code snippets. Additionally, code debugging and completion tasks trigger DFG retrieval in parallel to narrow the search space (for code debugging) or enable similarity matching between incomplete and complete code snippets (for code completion). \textbf{Step 3} utilizes LLMs to fuse the retrieved code snippets for code debugging, completion, or search, which further demonstrates the generality of our framework (see Section \ref{sub:code search}). In addition, due to the lack of dedicated code search benchmarks in HDL repositories, we build a new benchmark, HDLSearch, using an automated benchmarking build pipeline. Details of benchmark generation are presented in Section \ref{sub:benchmark}.

\subsection{Graph Database Preparation}
\label{sub:build graph database}

\label{sub:code search}
\label{sec:fr}

As shown in \textbf{Step 1} of Figure \ref{fig1}, the HDLxGraph RAG framework begins with an off-line graph database construction. 
Without losing representability, we focus on Verilog in our implementation. Please note that, although different HDLs have different syntactic properties, they share the same three-level structural abstraction as \texttt{module} $\rightarrow$ \texttt{block} $\rightarrow$ \texttt{signal} in Verilog. Specifically, we use an AST to support the \textbf{code graph view} that emphasizes multi-level structural relationships in HDL, and a DFG to facilitate the \textbf{hardware graph view} focusing on signal flow reflecting circuit topology, providing a comprehensive and tailored representation of the HDL repository. 
The AST graph incorporates node types such as \colorbox{mygray}{\texttt{MODULE}}, \colorbox{mygray}{\texttt{BLOCK}}, and \colorbox{mygray}{\texttt{SIGNAL}} connected through \colorbox{mygray}{\texttt{CONTAINS}} and \colorbox{mygray}{\texttt{INSTANTIATE}} edge types, whereas the DFG graph introduces \colorbox{mygray}{\texttt{TEMP}} nodes alongside \colorbox{mygray}{\texttt{SIGNAL}} nodes, connected via \colorbox{mygray}{\texttt{FLOWS\_TO}}, \colorbox{mygray}{\texttt{TRUE}}, \colorbox{mygray}{\texttt{FALSE}}, and \colorbox{mygray}{\texttt{COND}} edges. Figure~\ref{fig3:Visualization} demonstrates a toy example of Verilog source code, its corresponding nodes and edges, and the constructed graph database.
When constructing the entire graph database, we perform the following three sub-steps:


\begin{figure}[b]
     \centering
\includegraphics[width=1.0\linewidth]{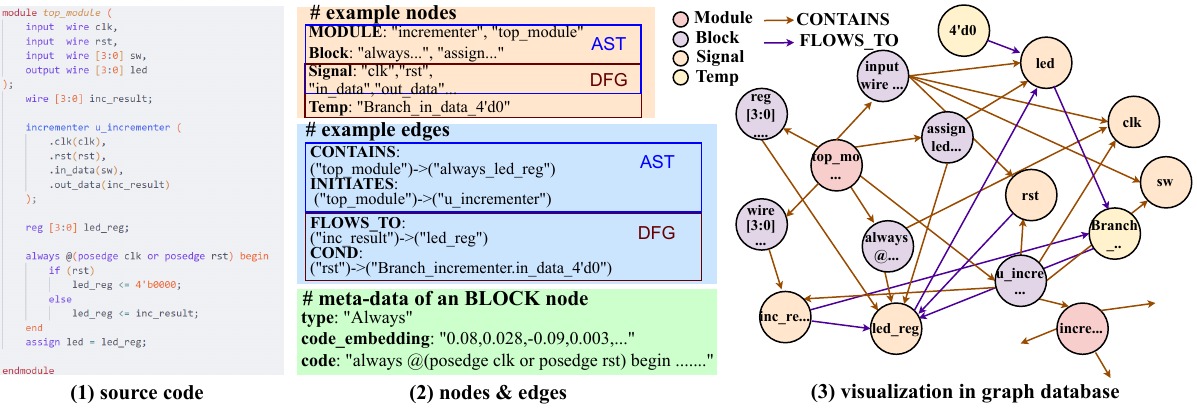} 
    \vspace{-2em}
    \caption{Visualization of an example in the graph database.}
    \label{fig3:Visualization}
\end{figure}

\textbf{1) Parsing.} The graph database construction begins with analyzing individual HDL file in the repository using a Verilator-based \cite{Snyder_Verilator} AST and DFG parser. For AST parsing, we extract the cross-level dependancy information of \colorbox{mygray}{\texttt{MODULE}}, \colorbox{mygray}{\texttt{BLOCK}}, and \colorbox{mygray}{\texttt{SIGNAL}} from each Verilog file to represent the fine-grained hierarchical code structure. Note that block level (always, assign, initial) here represent behavioral abstraction at the register-transfer level, defining concurrent hardware operations. Concurrently, we generate the hardware signal flow for DFG parsing, which characterizes the transmission and interaction between signals. The DFG graph incorporates both the signal directions and the dependency relationships between signals, reflecting the functionality and processing data flow of a circuit. This multi-granularity representation enables our database to store both the code structure and the hardware behavior of a single HDL file, thereby facilitating a comprehensive graph abstraction of the HDL, as shown in Figure \ref{fig3:Visualization} (2) and (3).

\textbf{2) Meta-data Generation.} After the parsing of graph data, we generate embeddings for nodes (both \colorbox{mygray}{\texttt{MODULE}} and \colorbox{mygray}{\texttt{BLOCK}}) via code encoding to facilitate semantic search. These embeddings and parser-extracted attributes (e.g., block type, code) are stored as node meta-data, as illustrated by the meta-data on the nodes and edge types in Figure~\ref{fig3:Visualization}~(3). We use CodeT5+ \cite{wang-etal-2023-codet5}, a SOTA code LLM model, to directly generate the embeddings for our parsers, thereby eliminating the need for intermediate description generation.



\textbf{3) Cross-file Relationship Construction.} Finally, we address the absense of cross-file relationship, which is the module \colorbox{mygray}{\texttt{INSTANTIATE}} relationships. We search for the module node with same name recorded in the meta-info instance block to establish the cross-file and cross-module relationships.




The developed graph database provides a multi-level code exploration spanning from module-level abstractions to signal-level implementations, thereby positioning our HDL graph database as a extensible framework for multiple downstream tasks thanks to the modular fashion of database schema management.

\subsection{Multi-level Retrieval for Downstream Task Completion}
\label{sub:code search}
\label{sec:fr}
\textbf{Multiple-level Retrieval.}
In real-world hardware project issues, user queries always contain rich contextual cues, such as module names, functional descriptions, and sometimes brief code snippets, offering hints for retrieval. Therefore, user queries can be used to extract multi-level structural information and then guide the following multiple-level AST retrieval.
In addition, signal-level flow through DFG retrieval is adopted for code completion and debugging tasks. 

\begin{figure}[t]
     \centering
\includegraphics[width=1.0\linewidth]{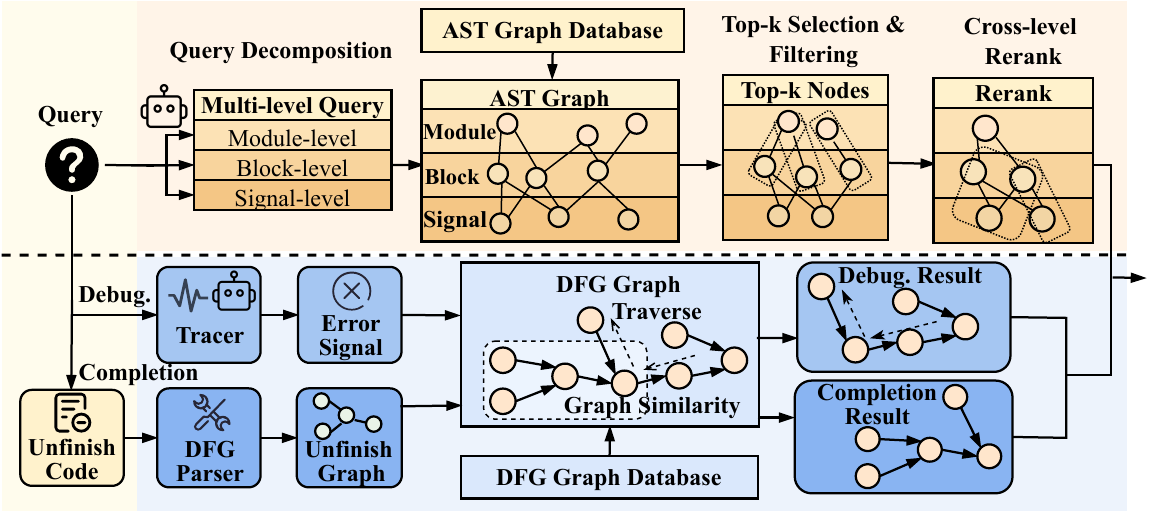} 
    \vspace{-2em}
    \caption{Flow of multi-level retrieval containing AST and DFG retrieval.}
    \label{fig4:Graph Retrieval}
\end{figure}

\textbf{AST Retrieval.} HDLxGraph constructs a hierarchical representation of HDL codebases through an AST-based graph, enabling multi-level HDL retrieval as depicted in Figure \ref{fig4:Graph Retrieval}. For AST retrieval, we follow three sub-steps:

\textbf{1) Query Decomposition.} HDLxGraph employs an LLM, called \textit{Decomposer}, to decompose the original query into three abstraction levels: module, block, and signal, thereby extracting structural information. It supports intricate queries from various downstream tasks such as: \texttt{`Find some certain blocks under a certain module'} in Search, or \texttt{`Some functions in some certain modules have led to the following errors ...'} Therefore, we obtain multi-level queries which have captured inherent structural information in the original query.

\textbf{2) Top-k Selection and Filtering.}  Leveraging Verilog's inherent three-level abstraction \texttt{module} $\rightarrow$ \texttt{block} $\rightarrow$ \texttt{signal}, HDLxGraph first retrieves top-k candidate modules and blocks which have the highest similarity scores with the decomposed query in corresponding levels based on semantic matching, then filters valid module-block pairs through containment relationships.
To facilitate precise code retrieval in different levels, a suite of retrieval APIs is introduced. Since we select Neo4j as the graph database, the query APIs are written in Cypher to interact with the database. 

\textbf{3) Cross-level Rerank.} Finally, we rerank results by averaging the similarity scores of module-block pairs containing each signal, since signal-level queries lack sufficient context for direct scoring. Therefore, HDLxGraph extracts all filtered module-block pairs that contain the signal and computes their average similarity scores as the signal-level similarity score so that we prioritize signal with the highest similarity score to be the retrieved signal. 
This hierarchical approach ensures fine-grained retrieval of HDL's structural information across multiple abstraction layers while maintaining compatibility with similarity-based semantic analysis, balancing precision and scalability in hardware database exploration.

\textbf{DFG Retrieval.} The DFG captures signal-level variables and their relationships, making it especially useful for signal-sensitive downstream tasks. In this work, we leverage it to enhance code completion and debugging, as shown in Figure \ref{fig1}. There are two primary operations for DFG retrieval of different tasks, which are \textit{Signal Traverse} and \textit{Similarity-based Extract}:

\textbf{1) Debugging.} For debugging tasks, if a signal mismatch is detected, the debugging process can iteratively traverse the DFG upstream with \textit{Signal Traverse} operation from the error signal, inspecting each node (e.g., operators, multiplexers, or instance outputs) to identify where the dataflow diverges from expected behavior. This approach guides LLM debugging by focusing only on the subgraph directly influencing the problematic signal, filtering out irrelevant code regions.  By extracting the immediate upstream nodes and their associated code blocks, the system can find a concise, context-rich error candidate set. 

\textbf{2) Completion.} While we want to retrieve the similar code with the unfinished code, some reference code may look different but still have similar functionality because the hardware (i.e., dataflow) described is very similar. Graph embedding offers a viable approach for Verilog code completion by translating code’s structural and semantic relationships into a unified mathematical framework. By leveraging GraphSAGE \cite{hamilton2018inductiverepresentationlearninglarge}, these graphs are compressed into low-dimensional vector representations that preserve contextual patterns, such as recurring HDL constructs (e.g., finite state machines, pipelined operations) or common coding idioms (e.g., non-blocking assignments in clock-driven blocks).
This allows real-time retrieval of relevant patterns from large codebases while adhering to Verilog-specific constraints.


\subsection{HDLSearch Benchmark}
\label{sub:benchmark}

 \begin{figure}[t]
     \centering
\includegraphics[width=0.86\linewidth]{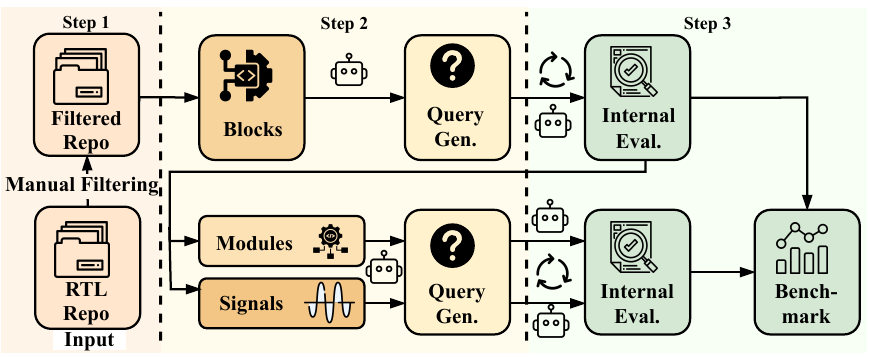} 
    \vspace{-1em}
    \caption{HDLSearch benchmark generation flow.}
    \label{fig:benchmark}
\end{figure}

While existing HDL code benchmarks inherently require HDL code search capabilities to evaluate task performance, they were not explicitly designed to assess these search capabilities in isolation. 
However, manually creating an expert-annotated benchmark is time-consuming and labor-intensive. 
Therefore, we propose an automated benchmark build pipeline that can automatically analyze, divide, and pair documentation texts and code snippets and construct a benchmark, dubbed HDLSearch. As shown in Figure~\ref{fig:benchmark}, the benchmark pipeline contains three main steps:

\textbf{1) Manual Filtering.} Our corpus originates from RTL-Repo \cite{allam2024rtlrepobenchmarkevaluatingllms}, a collection of publicly accessible GitHub repositories specializing in HDLs. Some HDL projects lack structured documentation and standardized code organization. Therefore, we first implement a manual filtering and select 10 representative repositories at different difficulty levels, ranging from educational FPGA projects, interconnection protocols to commercial CPUs.  

\textbf{2) Query Generation.} Adopting a hierarchical framework where \textbf{block serves as the fundamental level}, we implement a multi-stage generation process. Initial functional block descriptions are first generated, then systematically propagated through two parallel pathways, which are 1) \textbf{Signal-level annotation:} 
through contextual information, the semantics of a functional block can be inherited by its associated signals, thereby effectively annotating these signals with specific functionalities,
and 2) \textbf{Module-level abstraction} by designing a set of explicit and tailored prompts for the LLMs, we enable it to analyze and summarize the interactions among individual functional blocks as a module-level description.
This dual-path flow ensures consistent semantic alignment between fine-grained signal behaviors and coarse-grained module operations. With all descriptions finished, repo-specific information such as module and signal's names are removed to generate a relatively ambiguous query.

\textbf{3) Benchmark Refinement.} 
To further ensure benchmark validity, we employ an iterative refinement process using templated instructions. Through multiple rounds of evaluation and regeneration, we gradually remove unsuitable queries and align the LLM-generated query outputs with practical requirements till it reaches the defined termination count $K$. 


%% file: Sections/4-Evaluation.tex
\section{Experiments}
\label{sec:ev}

\subsection{Experimental Configuration and Platforms}

We evaluate~\name~from three perspectives: 1) effectiveness over traditional retrieval methods across code search, debugging, and completion; 2) comparison with SOTA software-code Graph RAG baselines; and 3) ablation of key components. We test {\name} with three LLMs of varying scales: Claude-3.5-Sonnet\cite{Claude-3.5} (large, strong coding ability), Qwen2.5-Coder-7B~\cite{hui2024qwen25codertechnicalreport} (medium, coding-oriented model), and LLaMA-3.1~\cite{grattafiori2024llama3herdmodels} (small, general-purpose). All experiments use top-p = 1.0, temperature = 0.7, and are conducted on a 2×A6000 Linux server with 10 independent runs per task for statistical robustness.

\subsection{Code Semantic Search}
\textbf{Benchmark:} Considering the absence of benchmarks for HDL-specific code search, we use our proposed HDLSearch (see Section~\ref{sub:benchmark}) as the benchmark with termination count $K=7$ when generation. The generated benchmark comprises 50 module-level, 100 block-level, and 200 signal-level queries, with 6,300 code blocks from 10 repositories serving as distractor, i.e., retrieval scope. The evaluation focuses on block-level retrieval, which serves as a fundamental step to other downstream tasks such as debugging and completion.

\textbf{Metric:} We adopt the widely-used mean reciprocal rank (MRR) in RAG as the metric, which assesses whether the method returns correct results within the top-ranked outputs.


\textbf{Baselines:} We compare {\name} against two traditional similarity-based RAG methods, BM25 \cite{robertson2009probabilistic} and CodeT5+ embeddings \cite{wang-etal-2023-codet5}, as well as the SOTA software-code GraphRAG method proposed by Microsoft \cite{edge2024localglobalgraphrag}.

\begin{figure}[t]
     \centering
\includegraphics[width=0.75\linewidth]{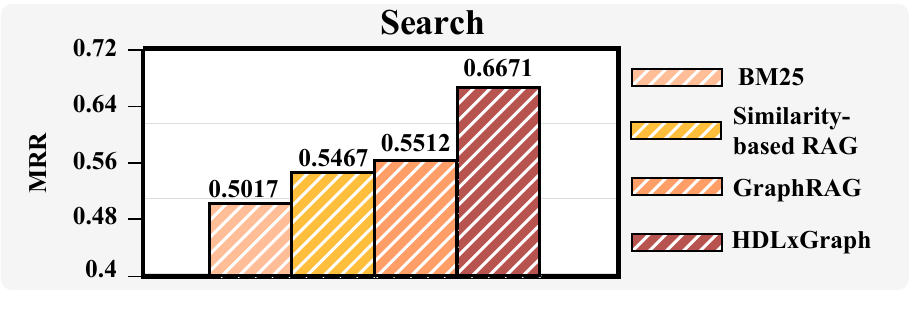} 
    \vspace{-1.3em}
    \caption{HDL search MRR comparison with baselines.}
    \label{ev:Search}
\end{figure}

\textbf{Evaluation Results:}
As shown in Figure~\ref{ev:Search}, {\name} achieves superior search performance across all 100 block-level queries than the similarity-based and software-code GraphRAG baselines. Notably, {\name} improves the average MRR by 12.04\% over the similarity-based baseline, whereas the sofware-code GraphRAG baseline achieves only a marginal improvement of 0.82\%. This highlights the necessity of incorporating HDL-inherent structural hierarchy (AST) and data dependencies (DFG) into RAG, enabling fine-grained retrieval and alignment with underlying hardware semantics compared to general Graph RAG methods.



\subsection{Code Debugging}
\label{sec:exp_code_debugging}
\textbf{Benchmark:} We choose the mor1kx repository~\cite{openrisc_mor1kx}, a repository-level debugging challenge in the SOTA LLM4SecHW~\cite{LLM4SecHW} benchmark, for our debugging evaluation. The mor1kx repository~\cite{openrisc_mor1kx} is an OpenRISC processor IP core with 5 git commit SHAs covering various debugging issues.

\textbf{Metric:} We choose ROUGE-N F1 score \cite{lin-2004-rouge} as the evaluation metric, which refers to the direct $N$-gram overlap between a prediction and a reference word considering precision and recall. The parameter $N$ can be set to 1, 2 and L, corresponding to matching at unigram, bigram, and longest common subsequence gram, respectively.


\textbf{Baselines:} We compare our framework with three RAG strategies: the accurate-RAG debugging strategy, denoted as ``Accurate-RAG", which relies on human effort to extract the exact buggy code segments to be modified as a theoretical top-tier RAG baseline, CodeT5+@\cite{wang-etal-2023-codet5}, denoted as ``Similarity-based RAG", which represents the conventional similarity-based RAG approach, and ``GraphRAG", which represents Microsoft's software-code GraphRAG.

 \begin{figure}[t]
     \centering
\includegraphics[width=1.0\linewidth]{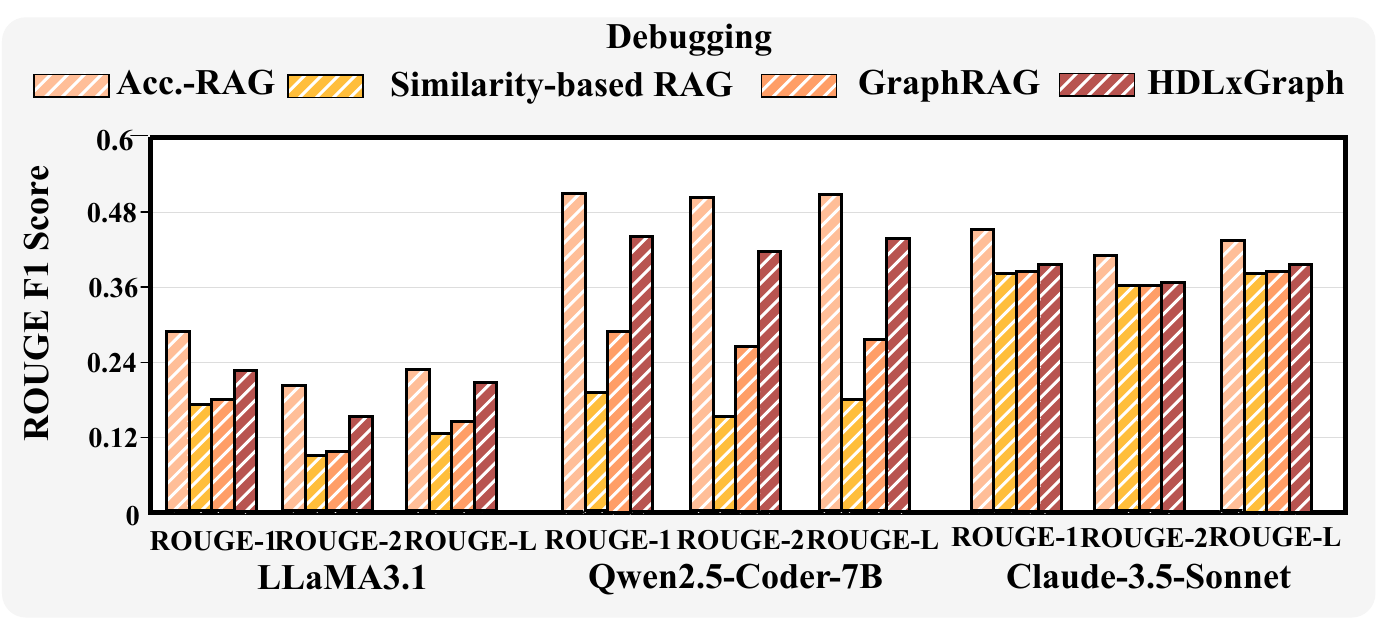} 
    \vspace{-2.2em}
    \caption{HDL debugging comparison with baselines.}
    \label{ev:Debugging}
\end{figure}


\begin{figure}[b]
     \centering
\includegraphics[width=1.0\linewidth]{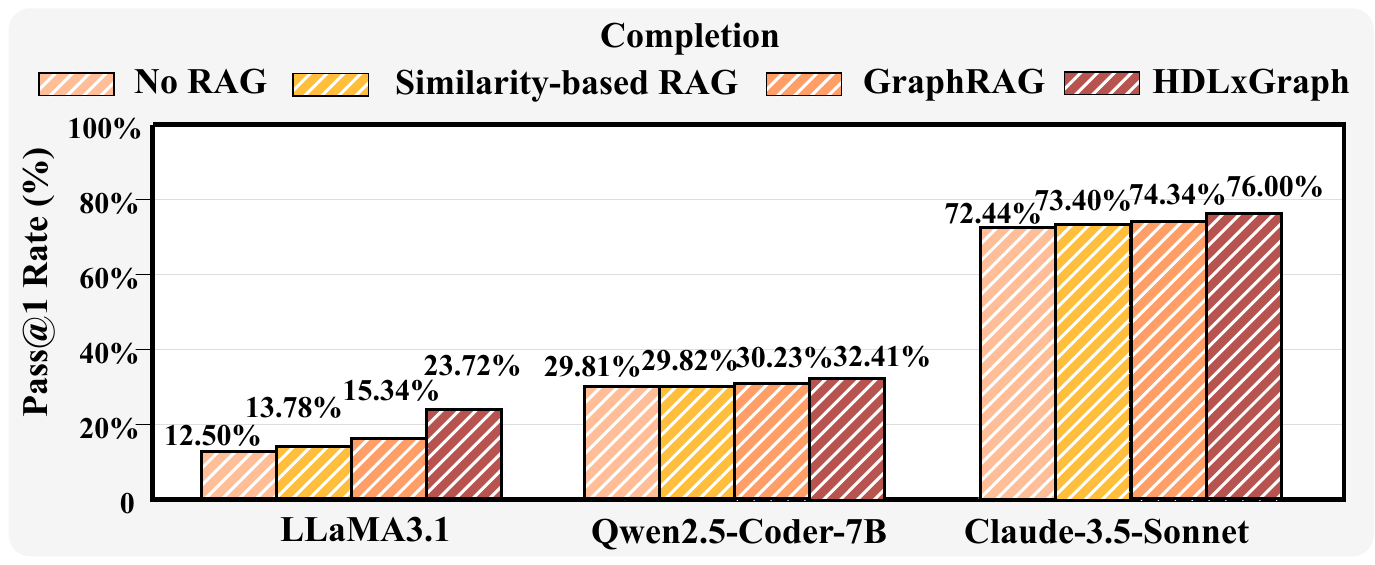} 
    \vspace{-2.3em}
    \caption{HDL completion Pass@1 comparison with baselines.}
    \label{ev:Completion}
\end{figure}

\textbf{Evaluation Results:}
As illustrated in Figure~\ref{ev:Debugging}, {\name} achieves consistently higher scores than both the similarity-based RAG and software-code Graph RAG baselines across all scenarios, with performance approaching that of ``Accurate-RAG".
Notably, when using less powerful LLaMA~3.1 and Qwen2.5-Coder-7B models, {\name} attains an average improvement of {8.18\%} in ROUGE-1, ROUGE-2, and ROUGE-L metrics over the best software Graph RAG baseline, demonstrating {\name}’s generalizability even under constrained hardware resources. 



\subsection{Code Completion}
\textbf{Benchmark:} We evaluate code completion capabilities using VerilogEval-Human v2~\cite{VerilogEval} with RTLLM~\cite{RTLLM} as a reference implementation. To avoid dataset contamination, we manually remove similar examples between the two datasets.

\textbf{Metric:} We apply Pass@k metrics \cite{ROME} to assess the generation pass rate.
$Pass@1$ is used in our experiment. 

\textbf{Baselines:} We compare {\name} against three baselines: direct LLM completion without RAG, the similarity-based RAG using CodeT5+, and the software-code GraphRAG.

\textbf{Evaluation Results:} As shown in Figure \ref{ev:Completion}, {\name} consistently improves $Pass@1$ accuracy by 3-10\% across various LLMs. While our evaluation framework operates at module granularity rather than full repository scope, we strategically employ the RTLLM\cite{RTLLM} codebase as a RAG corpus, thereby maintaining repository-level evaluation. 
The higher accuracy suggests {\name}'s generalizability across different abstraction levels, highlighting that structural code understanding significantly benefits completion tasks, even at sub-repository granularity. 

\subsection{Ablation Study}

\begin{table}[t]
    \centering
    \caption{Ablation Study on HDLxGraph Components}
    \vspace{-0.5em}
    \begin{tabular}{@{}lccc@{}}
        \toprule
        \textbf{Toolset Setting} & 
        \textbf{\makecell{Code Search\\MRR Score}} & 
        \textbf{\makecell{Code completion\\Pass@1 Rate(\%)}} & 
        \textbf{\makecell{Code debugging\\ROUGE L F1}} \\
        \midrule
        Ours (Full toolset) & 0.6671 & 23.72 & 0.205 \\
        \midrule
        w/o AST analysis & 0.1342 & 15.12 & 0.1438 \\
        w/o DFG behavior & 0.6312 & 22.54 & 0.1561 \\
        \bottomrule
    \end{tabular}
\end{table}

To evaluate the effectiveness of each component of HDLxGraph, we conduct an ablation study of AST and DFG separately. We use LLaMA3.1 as the model for these experiments.

\textbf{Evaluation Results:} We evaluate the results for the same three downstream tasks: 1) In code search, the AST hierarchical analysis plays a primary role, whereas DFG behavior detection has limited impact, as natural language queries rarely align directly with DFG-level semantics; 2) In code completion, similarly, the AST hierarchical analysis proves more effective due to its strong alignment with structural retrieval. In contrast, DFG has limited impact, as incomplete code often lacks sufficient signal-level context to construct meaningful dataflow; 3) In code debugging, both AST and DFG provide essential information, reflecting the human debugging process, which often involves tracing actual signal waveforms while simultaneously searching for relevant abstract code structures.



%% file: Sections/5-Conclusion.tex
\section{Conclusion and Future Work}
\label{sec:co}
We propose HDLxGraph, the first graph-enhanced RAG framework that combines AST-based structural matching with DFG-aware retrieval, for LLM-aided HDL tasks. Evaluations on search, debugging, and completion tasks show clear improvements over baselines, highlighting the effectiveness of {\name}. Future work will focus on developing adaptive traversal mechanisms for dynamic AST and DFG exploration, enabling more flexible interaction beyond fixed-function interfaces. We also aim to investigate multi-view HDL representations to bridge the gap between natural language and circuit semantics.